\documentclass[11pt]{article}

\usepackage{graphicx}
\usepackage{acronym}
\usepackage{amsmath} % allows eqref
\usepackage{multirow}
\usepackage{cite}
\usepackage[usenames,dvipsnames,svgnames,table]{xcolor}

\usepackage{flushend}

\newcommand{\eg}{\emph{e.g.}}

\newcommand{\ea}{\emph{et al.}}

\acrodef{BS}[BS]{base station}
\acrodef{DTX}[DTX]{discontinuous transmission}
\acrodef{PA}[PA]{power amplifier}
\acrodef{PC}[PC]{power control}
\acrodef{LTE}[LTE]{Long Term Evolution}
\acrodef{EARTH}[EARTH]{Energy Aware Radio and neTwork tecHnologies}
\acrodef{MIMO}[MIMO]{multiple-input multiple-output}
\acrodef{BA}[BA]{bandwidth adaptation}
\acrodef{RF}[RF]{radio frequency}
\acrodef{3GPP}[3GPP]{3rd Generation Partnership Project}
\acrodef{MS}[MS]{mains supply}
\acrodef{BB}[BB]{baseband}
\acrodef{DC}[DC]{direct-current}

\begin{document}

\title{A Parameterized Base Station Power Model}

\author{{Hauke~Holtkamp, Gunther Auer}\vspace{3mm}\\
DOCOMO Euro-Labs\\
D-80687 Munich, Germany \\Email:
\{holtkamp, auer\}@docomolab-euro.com
\and Vito~Giannini\vspace{2mm}\\
IMEC\\
Leuven, Belgium\\
vito@imec.be\\
\and Harald~Haas\vspace{2mm}\\
Institute for Digital Communications\\
Joint Research Institute for Signal and Image Processing\\
 The University of Edinburgh,
EH9 3JL, Edinburgh, UK\\ E-mail: h.haas@ed.ac.uk}

% \author{Author 1 Placeholder, Author 2 Placeholder \\Author 3 Placeholder, Author 4 Placeholder}

\maketitle

\begin{abstract}
Power models are needed to assess the power consumption of cellular \acp{BS} on an abstract level. Currently available models are either too simplified to cover necessary aspects or overly complex. We provide a parameterized linear power model which covers the individual aspects of a \ac{BS} which are relevant for a power consumption analysis, especially the transmission bandwidth and the number of radio chains. Details reflecting the underlying architecture are abstracted in favor of simplicity and applicability. We identify current power-saving techniques of cellular networks for which this model can be used. Furthermore, the parameter set of typical commercial \acp{BS} is provided and compared to the underlying complex model. The complex model is well approximated while only using a fraction of the input parameters.
\end{abstract}
% \begin{IEEEkeywords}
% energy efficiency, power model, earth, power consumption, mobile communications
% \end{IEEEkeywords}

% reset all acronyms in case they have been used in the abstract
\acresetall

\section{Introduction}
Recently, the power consumption of cellular networks has become a point of interest in research and even been taken into consideration for the standardization of future cellular networks like \ac{LTE}-Advanced~\cite{std:3gpp36927}. It was found in~\cite{fmbf1001} that in cellular networks the element which causes the largest share of overall consumption is the \ac{BS}. Numerous techniques have consequently been proposed by which the power consumption of \acp{BS} can be reduced~\cite{chp0801,cgb0401,hah1101,xylzcx1101,a1201,fmmjg1101}. Some of these techniques only consider transmission power while others take into consideration that the generation of the radio signal also consumes power in circuitry by employing power models. Such power models describe abstractly how much power a transmitter consumes and how this consumption depends on operating parameters. In the past, the modelling of \ac{BS} power consumption often had to be based on intuition until the first power models were published~\cite{djm1201,arfb1001,aggsoigdb1101,aggsoisgd1101,ddgfahwsr1201}. Simple models like~\cite{djm1201,arfb1001} allow computing the power consumption of a \ac{BS} for specific configurations. In contrast, the non-linear complex model described by Desset \ea~\cite{ddgfahwsr1201} is derived from the combination of each of a \ac{BS}' subcomponents. This allows inspecting the power consumption to such detailed level as the effect of giga operations per second or transistor gate lengths, but is unwieldy to apply. In this paper, we extend the work in~\cite{aggsoigdb1101,aggsoisgd1101} by maintaining simplicity while integrating two relevant operating variables into the model, namely the \ac{PA}'s output range and the transmission bandwidth. The proposed model allows assessing the power consumption of all techniques that are currently employed to reduce the power consumption of \ac{BS} while conserving simplicity. %Like the models in~\cite{aggsoigdb1101,aggsoisgd1101,ddgfahwsr1201}, this model is based on results obtained within the EARTH\footnote{EU funded research project EARTH (Energy Aware Radio and neTwork tecHnologies), FP7-ICT-2009-4-247733-EARTH, Jan.~2010 to June~2012. https://www.ict-earth.eu} project.

The scope of the model is described in Section~\ref{scope}. The model is subsequently presented in Section~\ref{model}. In Section~\ref{results}, it is discussed and compared to the complex model. The paper is concluded in Section~\ref{conclusion}. 
% Removed refs during review:
% lcvc1101,mhl0801,hafm1201,vh1101,abh1101,acc1201,d1201

% REFS:
% deruyck power model \cite{djm1201}
% desset ddgfahwsr1201
% arnold arfb1001
% commmag \cite{aggsoigdb1101}
% vtc11 \cite{aggsoisgd1101}
% 
% Power control
% - holtkamp, \cite{hah1101}
% lopez, \cite{lcvc1101}
% miao \cite{mhl0801}
% 
% Antenna Adaptation
% - xu, \cite{xylzcx1101}
% hedayati, \cite{hafm1201}
% 
% Bandwidth Adaptation
% - ambrosy, \cite{a1201}
% videv \cite{vh1101}
% 
% DTX/Sleep
% - frenger, \cite{fmmjg1101}
% ashraf, \cite{abh1101}
% abdallah, \cite{acc1201}
% de domenico \cite{d1201}
% 
% Remote Radio Head functionality
% - no citation
% 
% HetNets
% - fehske, \cite{frf0901}
% claussen \cite{chp0801}
% 
% Hardware design (cooling, conversion)
% cui \cite{cgb0401}
\section{Power Model Scope}
\label{scope}
Power saving techniques in literature can be generally divided into design changes and operating approaches. Design changes affect the layout of the network or the hardware architecture. For example, in \cite{frf0901} it is proposed that the use of heterogeneous networks will positively affect the network power consumption under certain conditions. Cui \ea~\cite{cgb0401} show how adjusting the number of antennas affects power consumption. In contrast to design changes, operating approaches manipulate the functionality of a \ac{BS} during operation. Here, proposed techniques are the reduction of transmission power~\cite{hah1101}, the deactivation of unneeded antennas~\cite{xylzcx1101}, the adaptation of the transmission bandwidth~\cite{a1201} and the use of low power consumption sleep modes of varied durations~\cite{fmmjg1101}.
% Refs removed during review: chp0801, lcvc1101,mhl0801,hafm1201,vh1101,abh1101,acc1201,d1201

The model presented in this paper encompasses all of these approaches to allow for a direct comparison while abstracting parameters which can either be assumed to be constant or have been shown to have little effect in the studied scenarios, such as modulation and coding settings, equipment manufacturing details and leakage powers~\cite{ddgfahwsr1201}. To this extent, the following is covered in the proposed model:
\begin{itemize}
 \item The different \ac{BS} types of a heterogeneous network are modelled by applying different parameter sets to the same model equations. 
 \item The number of transmission antennas and radio chains affects consumption during design and operation.
 \item The same holds true for transmission power, which affects the design indirectly by choice of a suitable power amplifier as well as the operation directly. 
 \item Also, transmission bandwidth and sleep modes are modelled in their effect on \ac{BS} power consumption.
\end{itemize}
\section{Base Station Power Model}
\label{model}
\begin{figure}
\centering
\includegraphics[width=0.4\textwidth]{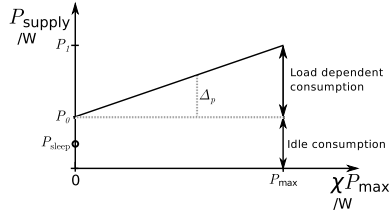}
 \caption{Load-dependent power model for an \ac{LTE} \ac{BS}.}
\label{fig:loaddeppm}
\end{figure}

It was found in~\cite{aggsoigdb1101} that the supply power consumption of a \ac{BS} can be approximated as an affine function of transmission power. In other words, the consumption can be represented by a static (load-independent) share, $P_0$, with an added load-dependent share that increases linearly by a power gradient, $\Delta_{\mathrm{p}}$. The maximum supply power consumption, $P_1$, is reached when transmitting at maximum total transmission power, $P_{\mathrm{max}}$. Furthermore, a \ac{BS} may enter a sleep mode with lowered consumption, $P_{\mathrm{sleep}}$, when it is not transmitting. Fig.~\ref{fig:loaddeppm} shows an illustration. Total power consumption considering the number of sectors, $M_{\mathrm{sec}}$, is then formulated as
\begin{equation}
\label{eq:psupply}
 P_{\mathrm{supply}}(\chi) = 
  \begin{cases}
   M_{\mathrm{sec}}  \left( P_1 + \Delta_{\mathrm{p}} P_{\mathrm{max}} (\chi - 1) \right)			& \text{if } 0 < \chi \leq 1  \\
   M_{\mathrm{sec}} P_{\mathrm{sleep}}      	& \text{if } \chi = 0,
  \end{cases}
\end{equation}
where $P_1 =  P_0 + \Delta_{\mathrm{p}} P_{\mathrm{max}}$. The scaling parameter $\chi$ is the load share, where $\chi=1$ indicates a fully loaded system, \eg~transmitting at full power and full bandwidth, and $\chi=0$ indicates an idle system.
To further understand the contribution of different parameters on this basic model, we parameterize the maximum supply power consumption, $P_1$. We first establish how power consumption scales with the transmission bandwidth in Hz, $W$, the number of \ac{BS} radio chains/antennas, $D$, and the maximum transmission power in W, $P_{\mathrm{max}}$. This requires to consider the main units of a \ac{BS}: \ac{PA}, \ac{RF} small-signal transceiver, \ac{BB} engine, \ac{DC}-\ac{DC} converter, active cooling and \ac{MS}. The dependence of the \ac{BS} units on $W$, $D$ and $P_{\mathrm{max}}$ can be approximated as follows~\cite{ddgfahwsr1201}:
\begin{itemize}
 \item Both the power consumptions of \ac{BB} and \ac{RF}, $P_\mathrm{BB}$ and $P_\mathrm{RF}$, respectively, scale linearly with bandwidth, $W$, in and the number of \ac{BS} antennas $D$. For some basic consumptions, $P_\mathrm{BB}^\prime$ and $P_\mathrm{RF}^\prime$, we thus define
\begin{equation}
P_\mathrm{BB} = D \frac{W}{10~\mathrm{MHz}} P_\mathrm{BB}^\prime
\end{equation}
and
\begin{equation}
 P_\mathrm{RF} = D \frac{W}{10~\mathrm{MHz}} P_\mathrm{RF}^\prime.
\end{equation}
 \item The \ac{PA} power consumption $P_{\mathrm{PA}}$ depends on the maximum transmission power per antenna $P_{\mathrm{max}}/D$ and the \ac{PA} efficiency $\eta_\mathrm{PA}$. Also, possible feeder cable losses, $\sigma_\mathrm{feed}$, have to be accounted for:
\begin{equation}
\label{ppa}
 P_{\mathrm{PA}} = \frac{P_{\mathrm{max}}}{D \eta_\mathrm{PA}(1{-}\sigma_\mathrm{feed})}.
\end{equation}
 \item Losses incurred by \ac{DC}-\ac{DC} conversion, \ac{MS} and active cooling scale linearly with the power consumption of other components and may be approximated by the loss factors $\sigma_\mathrm{DC}$, $\sigma_\mathrm{MS}$, and $\sigma_\mathrm{cool}$, respectively. These losses are included in the model as losses of a total according to~\cite{aggsoigdb1101}. Active cooling is typically only applied in Macro type \acp{BS}.
\end{itemize}

These assumptions are combined to calculate the maximum power consumption of a \ac{BS} sector,
\begin{subequations}
\begin{align}
 P_1 	&= \frac{ P_\mathrm{BB} + P_\mathrm{RF}  + P_{\mathrm{PA}}}
      {\displaystyle (1{-}\sigma_\mathrm{DC})(1{-}\sigma_\mathrm{MS})(1{-}\sigma_\mathrm{cool}) } \\
	&= \frac{ D \frac{W}{10~\mathrm{MHz}} \left( P_\mathrm{BB}^\prime + P_\mathrm{RF}^\prime   \right) + \frac{P_{\mathrm{max}}}{D \eta_\mathrm{PA}(1{-}\sigma_\mathrm{feed})} }
      {\displaystyle (1{-}\sigma_\mathrm{DC})(1{-}\sigma_\mathrm{MS})(1{-}\sigma_\mathrm{cool}) }. 
\end{align}
\label{p1}
\end{subequations}
An important characteristic of a \ac{PA} is that operation at lower transmit powers reduces the efficiency of the \ac{PA} and that, consequently, power consumption is not a linear function of the \ac{PA} output power. This is resolved by taking into account the ratio of maximum transmission power of the \ac{PA} from the data sheet, $P_{\mathrm{PA,limit}}$ to the maximum transmission power of the \ac{PA} during operation $\frac{P_{\mathrm{max}}}{D}$. The current transmission power can be adjusted by adapting the \ac{DC} supply voltage, which impacts the offset power of the \ac{PA}. The efficiency is assumed to decrease by a factor of $\gamma$ for each halving of the transmission power. The efficiency is thus maximal when $P_{\mathrm{max}}=P_{\mathrm{PA,limit}}$ in single antenna transmission and was heuristically found to be well-described by
\begin{equation}
 \eta_\mathrm{PA}= \eta_\mathrm{PA,max}\left[ 1-\gamma \log_2 \left( \frac{P_{\mathrm{PA,limit}}}{P_{\mathrm{max}} / D} \right)  \right],
\end{equation}
where $\eta_\mathrm{PA,max}$ is the maximum \ac{PA} efficiency.

The reduction of power consumption during sleep modes is achieved by powering off \acp{PA} and reduced computations necessary in the \ac{BB} engine. For simplicity, we only model the dependence on $D$ as each \ac{PA} is powered off.  Thus, $P_{\mathrm{sleep}}$, is approximated as
\begin{equation}
 P_{\mathrm{sleep}} = D P_{\mathrm{sleep},0},
\end{equation}
where $P_{\mathrm{sleep},0}$ is a reference value for the single antenna \ac{BS} chosen such that $P_{\mathrm{sleep}}$ matches the complex model value for two antennas.
\section{Results and Discussion}
\label{results}
\begin{figure}
\centering
\includegraphics[width=0.4\textwidth]{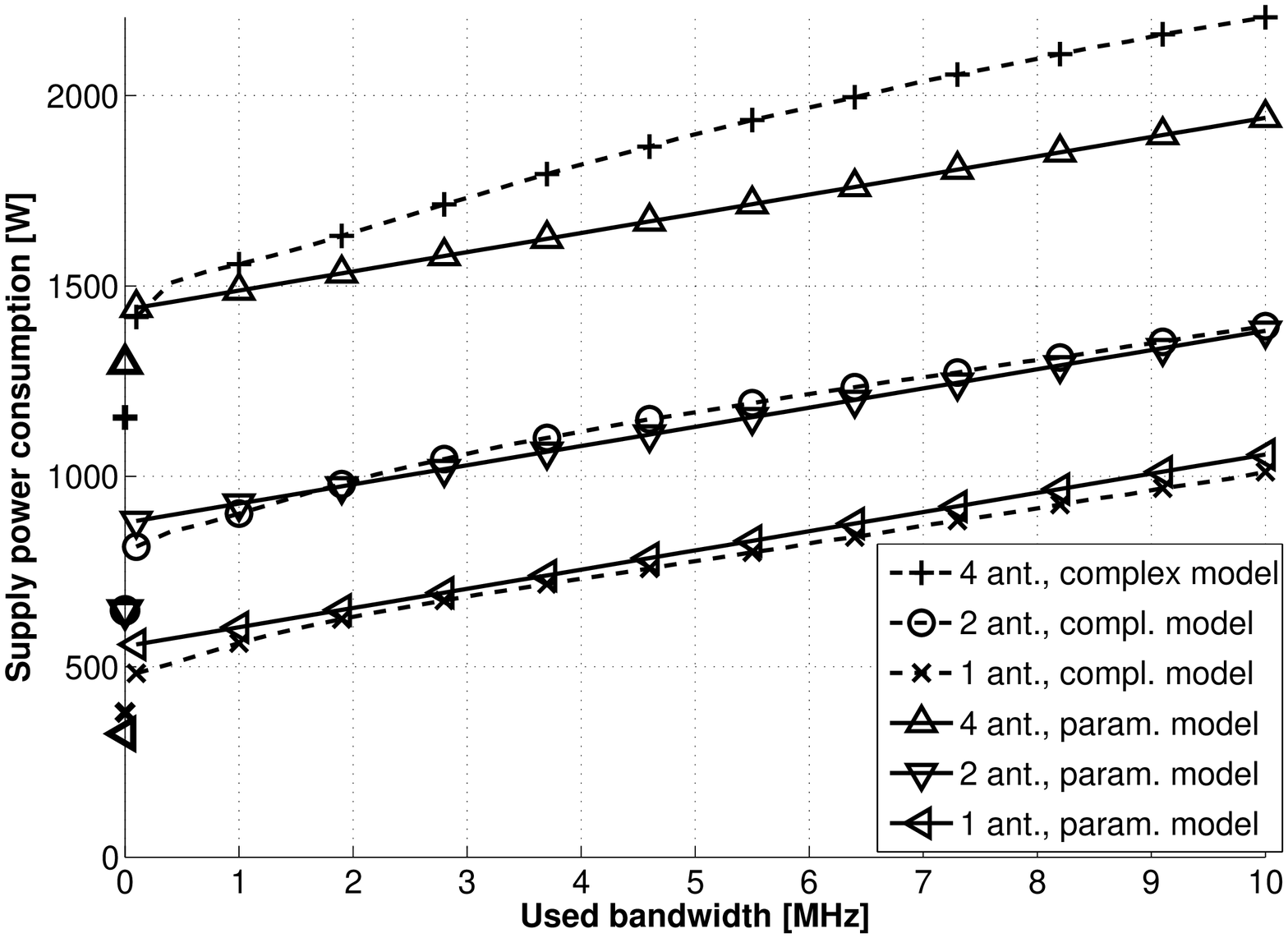}
 \caption{Comparison of the parameterized with the complex model~\cite{ddgfahwsr1201} power models for the Macro \ac{BS} type with 40~W transmission power.}
\label{fig:comparisonMacro}
\end{figure}

\begin{figure}
\centering
\includegraphics[width=0.4\textwidth]{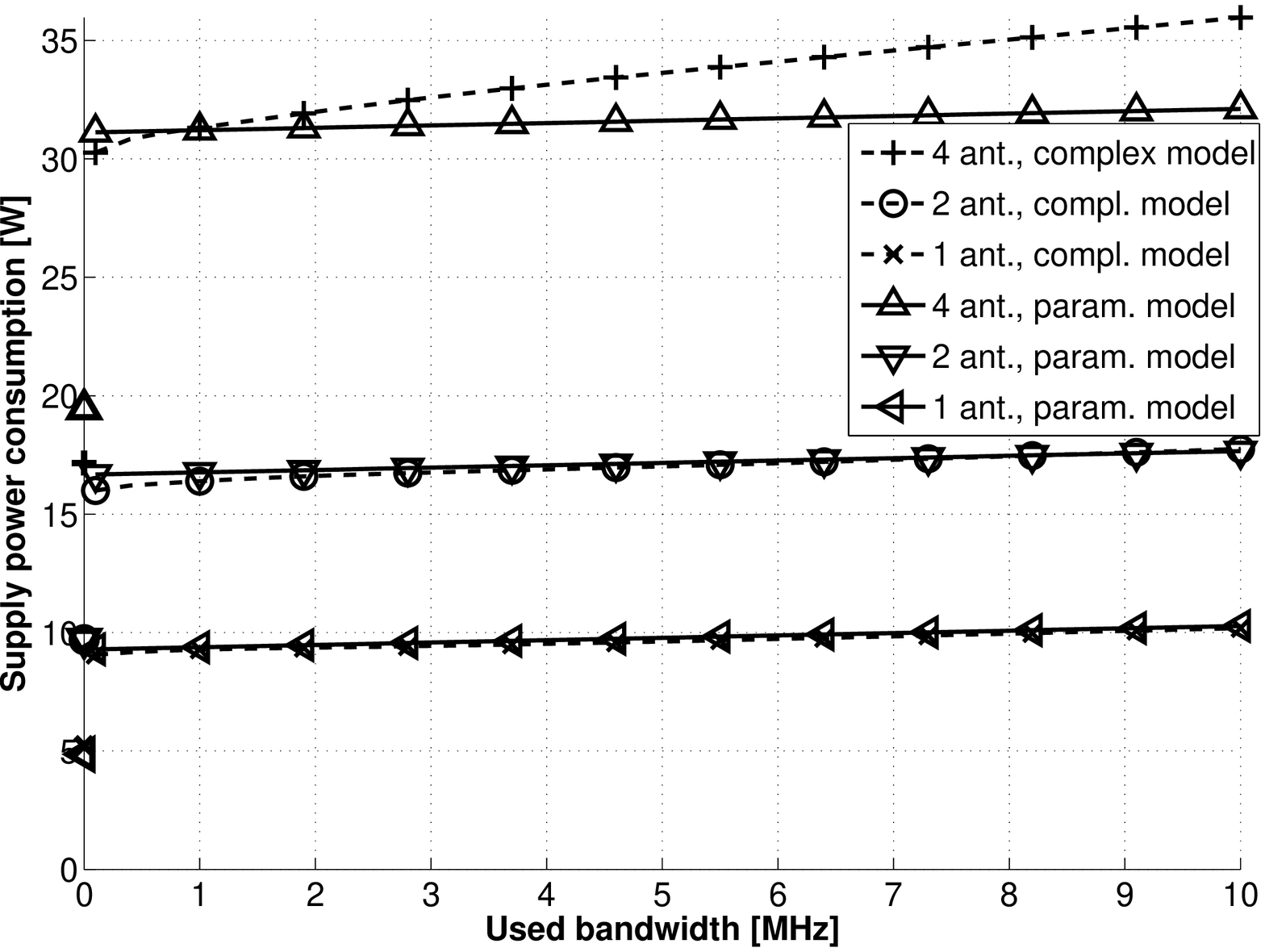}
 \caption{Comparison of the parameterized with the complex model~\cite{ddgfahwsr1201} power models for the Pico \ac{BS} type with 0.25~W transmission power.}
\label{fig:comparisonPico}
\end{figure}

\begin{figure}
\centering
\includegraphics[width=0.4\textwidth]{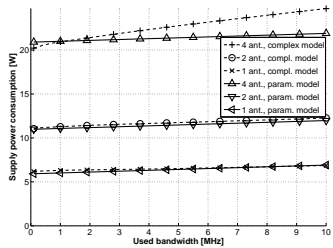}
 \caption{Comparison of the parameterized with the complex model~\cite{ddgfahwsr1201} power models for the Femto \ac{BS} type with 0.1~W transmission power.}
\label{fig:comparisonFemto}
\end{figure}

The parameterized power model is applied to approximate the consumption of the Macro, Pico and Femto \acp{BS} which are described in the complex model. Parameters are chosen where possible according to~\cite{aggsoigdb1101}, such as losses, efficiencies and power limits. The remaining parameters are adapted such that a closer match to the complex model could be achieved. The resulting parameter breakdown is provided in Table~\ref{tab:powerbreakdownmaximumload}. The proposed and the complex power models are compared for a bandwidth sweep with a varying number of transmit antennas in Fig.~\ref{fig:comparisonMacro}, Fig.~\ref{fig:comparisonPico}, and Fig.~\ref{fig:comparisonFemto}, for a Macro, a Pico and a Femto station, respectively. Although two parameters, the bandwidth and the number of \ac{BS} antennas, are varied, the parameterized model can be seen to closely approximate the complex model for all \ac{BS} types. The largest deviation of the parameterized model from the complex model occurs when modeling four transmit antennas. This is caused by the fact that the parameterized model considers a constant slope, $\Delta_{\mathrm{p}}$, which is independent of $D$. In contrast, the \ac{PA} efficiency in the complex model decreases with rising $D$, leading to an increasing slope which can not be matched by a constant slope. This deviation is a trade-off between simplicity and model accuracy. 

In addition to providing a solid reference, the model and the parameters can provide a basis for exploration. Individual parameters can be changed to observe the resulting variation in power consumption. With regard to the number of antennas, the parameterized model can only be verified up to four antennas, which is the extent of the complex model. Extending the system bandwidth, for example to 20~MHz, is expected to increase the \ac{BB} and \ac{RF} power consumption. The other parameters such as the transmission power and losses are expected to remain unaffected by different system bandwidths. Adapting the design maximum transmission power, $P_{\mathrm{max}}$, affects the \ac{PA} efficiencies, which decrease with $P_{\mathrm{max}}$.%
\begin{table*}[t]
\centering
\begin{tabular}{|c|c|c|c|c|c|c|c|c|c|c|c|c|c|c|c|}
\hline
\ac{BS} type 	& $P_{\mathrm{PA,limit}}$	& $\eta_\mathrm{PA,max}$	& $\gamma$	& $P_\mathrm{BB}^\prime$	& $P_\mathrm{RF}^\prime$	& $\sigma_\mathrm{feed}$	& $\sigma_\mathrm{DC}$	& $\sigma_\mathrm{cool}$	& $\sigma_\mathrm{MS}$ 	& $M_{\mathrm{Sec}}$ 	& $P_{\mathrm{max}}$  	& $P_{1}$  	 	& $\Delta_{\mathrm{p}}$ & $P_{\mathrm{sleep},0}$\\
		& /W			&				&		&	/W		&	/W		&				&			&				&			&			&		/W	&	/W	& *10\,MHz		& /W\\
\hline
Macro		 			& 	80.00	& 	0.36& 	0.15& 	29.4& 	12.9& 	0.5& 	0.075& 	0.1& 	0.09& 	3& 	40.00& 	460.4 & 		4.2 &	324.0 \\
\hline
Pico		 			&	0.25& 	0.08& 	0.20& 	4.0& 	1.2& 	0.0& 	0.09& 	0.0& 	0.11& 	1& 	0.25& 	17.4& 	4.0& 	4.9\\
\hline
Femto		 			& 	0.10& 	0.05& 	0.10& 	2.5& 	0.6& 	0.0& 	0.09& 	0.0& 	0.11& 	1& 	0.10& 	12.0& 	4.0& 	3.3\\
\hline
\end{tabular}
\caption{Parameter breakdown.}
\label{tab:powerbreakdownmaximumload}
\end{table*}%
\section{Conclusion and remarks}
\label{conclusion}
{In this paper, we have provided a parameterized power model which allows calculating the power consumption of a modern \ac{BS} based on important design and operation parameters. The model is much simpler and more applicable than the model it was derived from. A comparison of the parameterized model with the source model is provided.} %Note that this model is an instantaneous power consumption model. Considering the time it takes to switch between different operating states, for example, in and out of sleep mode, adds high complexity and is part of future work.

\bibliography{parametrizedPowermodel} 
\bibliographystyle{ieeetran}

\end{document}